# Effects of Adopting Ultra-Fast Charging Stations in the San Francisco Bay Area

Pouya Rezazadeh Kalehbasti[*,1], Yufei Miao[*,2], Gregory Andrew Forbes[*,3]

[*]Civil and Environmental Engineering, Stanford University
450 Serra Mall, Stanford, CA 94305
[1] `pouyar@stanford.edu`
[2] `miaoyf@stanford.edu`
[3] `gforbes@stanford.edu`

*Abstract*— **Ultra-Fast Charging (UFC) is a rising technology that can shorten the time of charging an Electric Vehicle (EV) from hours to minutes. However, the power consumption characteristics of UFC bring new challenges to the existing power system, and its pros and cons are yet to be studied. This project aims to set up a framework for studying the different aspects of substituting the normal non-residential EV chargers within the San Francisco Bay Area with Ultra-Fast Charging (UFC) stations. Three objectives were defined for three stakeholders involved in this simulation, namely: the EV user, the station owner, and the grid operator. The results show that, UFCs will significantly contribute to increase of peak load and power consumption during the peak demand period, which is an undesirable outcome from grid operation perspective. Total electricity and operations & maintenance (O&M) costs for station owner would increase subsequently, while this can be justified by analyzing the value of time (VOT) from an EV user's perspective. Additionally, peak-shaving using battery storage facilities is studied for complementing the applied technology change and mitigating the impacts of higher power consumption on the grid.**

## I. INTRODUCTION

Current EV charging station technology can take 1 to 20 hours to fully charge an EV's battery, depending on the station's power rating which normally ranges from 2 to 40 kW. However, typical vehicle users are accustomed to the short durations of refuelling gas-powered vehicles, resulting in negative consequences of long EV charging times. These consequences include users' insufficiently charging their EV's battery, changes in long-distance driving behaviour, and an unsatisfactory end-user experience.

One potential solution to the long charging times includes implementing UFC station technology. Proposed UFC stations operate between 350 kW to 1 MW of power to achieve charging times of less than 4 minutes per vehicle, which is comparable to the refuelling times of traditional gasoline stations [1]. With the proposed charging times, this can alter the charging behaviour of EVs and provide opportunities for EVs to be charged at a lower cost within the time-of-use pricing scheme in the San Francisco Bay area similar to the works by [2], [3].

In the current literature, there are a small number of large-scale studies such as [2] whereas most of the works focuses on a limited number of EVs, e.g. [4]. Additionally, most of such studies take many assumptions into consideration for modeling the traffic demand, such as [5]. In this project, following the method applied in [2], Virtual Aggregation Points (VAPs) were assumed that aggregated the power feed for a large quantity of EV power stations. Additionally, the actual arrival profiles were acquired from the same paper for the area of study.

This project aims to focus on the San Francisco Bay area while keeping the power grid configurations consistent over time. The goal is to understand how the technology upgrade and the shortened duration of charging times will affect the cost of electricity for EVs and the power consumption profile for the Bay area. The study will consider 0% to 100% of the level 2 charging stations in this region being replaced by ultra-fast chargers.

The Chevy Bolt is taken as the design target car with a 60 kWh battery [2], level 2 charging stations are modeled to take 20 hours to fully charge the target car, and UFC stations are modeled to take roughly 4 minutes fully charge the car. A time-of-use (TOU) pricing scheme is implemented to analyze the cost of EV charging based on PG&E's tables E-19 and E-20.

In this study, the UFC charging station implementation is considered from three points of view: (POVs): the EV user, the station owner, and the grid operator. The grid owner is considered to value minimizing the peak hour power consumption. The station owner considers minimizing the cost associated with fully charging the target vehicles. The EV power station user is also inclined to minimize the charging cost, in addition to minimizing the amount of time it takes to charge their EV.

## II. PROBLEM STATEMENT

The objective of this study is to understand how transitioning from level 2 charging stations to UFC charging stations will affect the three stakeholders involved in these transitions: the EV user, the station owner, and the grid operator.

The first part of this study involves examining the transition without affecting the power distribution of the system. The second part of this study involves examining this transition, while optimizing for each of the stakeholders' objectives by performing a power shaving analysis of the power distribution, supplemented by battery storage.

The scope of this problem will include 3500 EV charging plugs within the Bay area, where the scope boundary is shown in **Figure 1**. The analysis will be performed during the current year over the months of January (winter season) and August (summer season). The work presented in this paper will be performed with the use of the Python programming language.

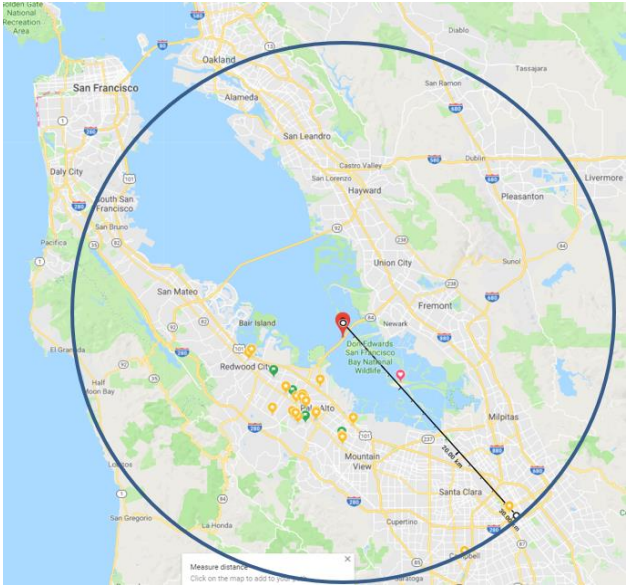

**Figure 1** – EV Charging Station Scope of Study

### III. METHODOLOGY

These problems of EV charging station transition and power shaving will be performed subsequent to one another and are not independent. The steps performed in this analysis to complete this study include:

A. Understanding the average usage of the Bay area's charging stations, throughout the day.

B. Modelling the loads of the EVs at the charging stations, throughout the day, in the Bay area.

C. Calculating the cost of charging an EV based on a TOU pricing scheme for both the winter and summer seasons.

D. Performing the above three modelling aspects from 0% to 100% UFC stations in the Bay area.

E. Model how the full charging station transition affects the three stakeholders associated in the scope.

F. Optimize for the objectives of each of the stakeholders by performing a power shaving analysis, coupled with battery storage, of the EV power distribution.

Regarding step (A) within the process, the EV arrival data for charging stations was taken from Kiliccote et. al. [2] This data included the charging station arrivals for the San Francisco Bay Area, Santa Rosa, Sacramento, and Los Angeles.

To find the proportion of the arrivals within the Bay area, the sum of the annual charging sessions in the Bay area was normalized by the sum of the total charging stations (Correction = 521,601 / (521,601 + 52,979) = 0.908). Additionally, the sum of plug-in vehicles at the end of 2017 in CA was divided by that of the end of 2013 to estimate the new values of daily, hourly, and monthly arrivals for the current study (Correction = 365,286 / 69,999 = 5.22).

**Figure 2** shows the modified yearly arrival profile over the course of a day for the Bay area of study.

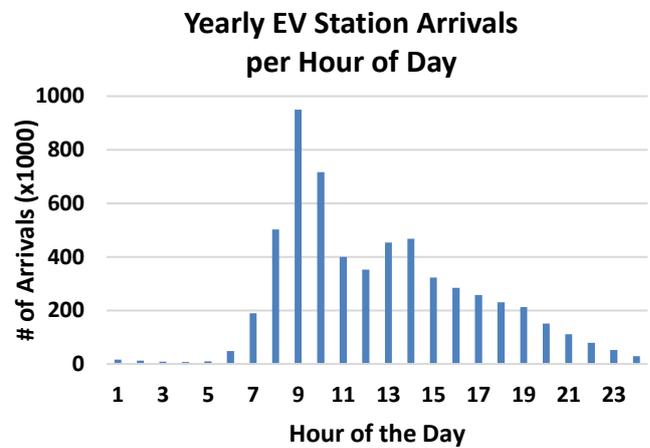

**Figure 2** – Modified arrivals' profile

Regarding step (B), the loads were modelled based on the amount of EV arrivals to a charging station within a specific period of time, the Chevy Bolt's battery capacity (60 kWh), and the power rating of the level 2 (3 kW) and UFC (900 kW) charging stations. The power rating for the UFC station was chosen as 900 kW, based on [2], which is a more conservative value. The power rating of the level 2 charging station was chosen after analysing power ratings of the charging stations in the Bay area, taken from the OpenChargeMap API. This distribution is shown in **Figure 3**, below. The legend shows the power ratings (in kW) associated with each color. Based on this distribution, most of the chargers (74%) were rated at 3 kW, thus it was selected as the representative level 2 charging station technology.

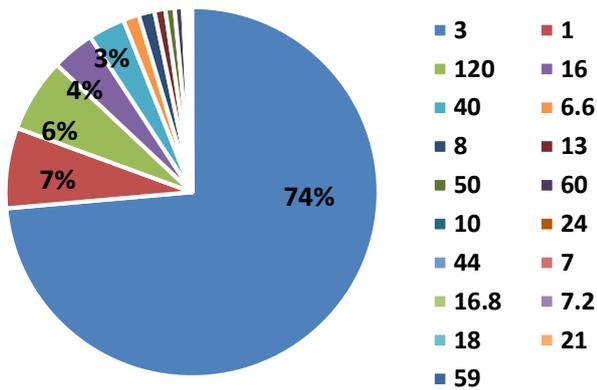

**Figure 3** – Station Power Rating Distribution in the Bay Area

It was assumed that the EVs arrived at the charging station completely uncharged and left the charging stations fully charged. The incoming load at a particular time is thus modelled as the product of the number of arrivals and the amount it takes to charge an EV based on the power station used.

For the case of the level 2 charging station, it takes 20 hours to fully charge a Chevy Bolt. For the case of the UFC station, it takes 4 minutes to fully charge a Chevy Bolt.

Regarding step (C), the cost at a particular hour is calculated based on annual O&M costs, uniform annual costs [6], and a TOU tariff scheme [7]. The O&M and annual costs are shown in **Table 1**, below.

| Charging Station Type | O&M Costs | Annual Costs |
|---|---|---|
| AC Level 2 | $200 | $592 |
| DC UFC | $2000 | $5904 |

**Table 1** – O&M and annual charging station costs

The TOU function charges for the energy used every 15 minutes and the maximum power demand over a period of a month. The two tariff schemes used for the analysis include PG&E's E-19 and E-20 tariff structures. For the level 2 charging stations, the E-19 (secondary user) TOU pricing scheme was used because level 2 charging stations met PG&E's requirements of medium demand. For the UFC charging stations being modeled, the E-20 (secondary user) cost function was used because UFC stations met the requirements of 1000 kW+ demands.

**Figure 4** shows the TOU energy cost function for the E-20 Tariff in the winter season. **Figure 5** shows the TOU cost function for the E-20 Tariff in the summer season. The TOU cost function for the E-19 Tariff presents an almost exact structure as that of E-20, except the cost are slightly higher.

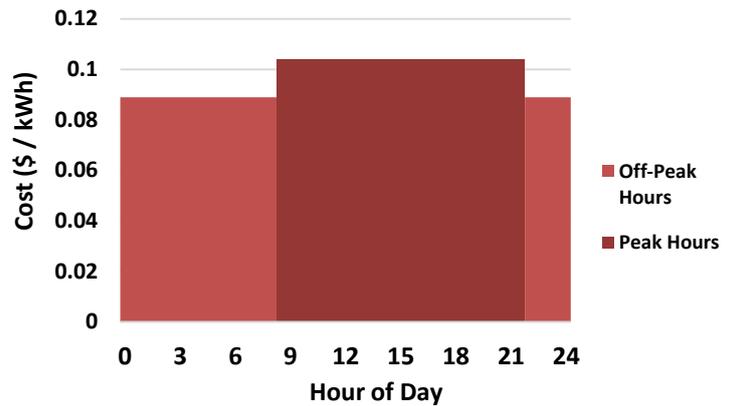

**Figure 4** – Winter TOU Cost Function

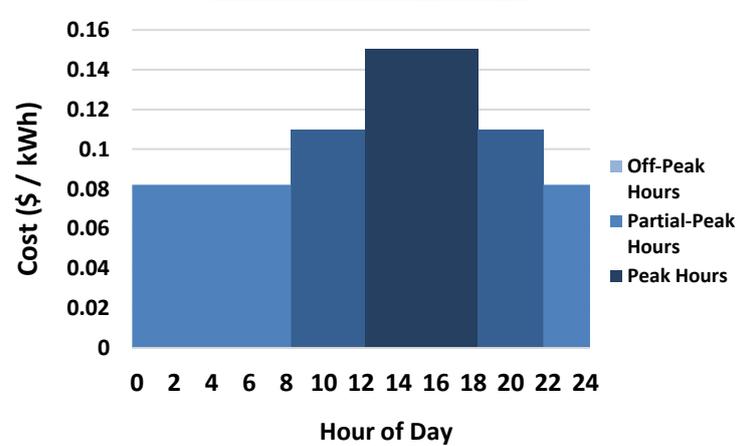

**Figure 5** – Summer TOU Cost Function

The demand charges involved in the TOU pricing schemes are calculated based on the highest power consumptions observed during partial-peak, peak, and all hours within a month's time span. These charges were incurred at the end of a month period, rather than 15 min. **Table 2** presents the demand charged for power in the E-19 and E-20 TOU pricing schemes for the winter and summer seasons.

|  | E-19 TOU ($ / kW) | E-20 TOU ($ / kW) |
|---|---|---|
| Peak Power – Summer | 19.02 | 19.65 |
| Partial-Peak Power – Summer | 5.23 | 5.40 |
| Max. Power – Summer | 17.87 | 17.74 |
| Partial-Peak Power – Winter | 0.05 | 0.12 |
| Max. Power - Winter | 17.87 | 17.74 |

**Table 2** – Power Demand Charges (E-19 & E-20)

Regarding step (D), this modelling approach was done by treating the Bay area as one node. First, it was assumed that all EV power stations within the Bay area consisted of 100% level 2 charging mechanisms. Then, incrementally the level 2 charging stations were converted into UFC stations until all 100% of the charging stations were UFC. This study was modelled for every increment.

For step (E), the transition from 100% level 2 charging stations to 100% UFC charging stations was examined from the POVs of the EV user, the station owner, and the grid operator. Each of their objectives, along with a description of the modelling that took place for each objective, is subsequently described.

For the EV user, the objective is to minimize both the combined time and cost associated with charging their EV. The cost is modelled through the TOU pricing schemes, as previously mentioned. Additionally, a value-of-time (VOT) metric scheme was used to quantify the cost associated with the time it takes to charge an EV. This cost associated with charging time was added to the TOU cost to model the total cost to the EV user. These additional (VOT) factors are shown in **Table 3**.

| VOT Type | VOT Quantity ($ / hr.) |
|---|---|
| No VOT | 0.00 |
| Tipping Point | 0.70 |
| Fed. Law Min. Wage | 7.25 |
| Uber Driver Avg. Income | 8.55 |

**Table 3** – Value-of-Time Weighting Factors

The station owner's objective is to minimize the cost of operating the EV charging stations. This cost is the same as that of the cost to the EV user. It was assumed that no profits were associated with this transaction because most of these charging stations are publicly operated.

The grid operator's objective includes minimizing the amount of power and the peak power observed during hours of peak demand in the Bay area. These power profiles were measured from 100% level 2 charging stations to 100% UFC charging stations.

Last, for step (F), once the modelling for all three objectives are performed and finished, a power shaving optimization coupled with battery storage is performed. This analysis optimizes for each of the users to minimize cost, power used during the peak demand period, and the peak power observed during the peak demand period.

## IV. UFC STATION POWER MODELING

At each UFC replacement increment (from 0 to 100), we split the arrivals per 4 minutes between UFC charging and AC charging according to the aforementioned arrival profile. The power profile for every 4 minutes is then generated for each charging technology according to its power rating and time needed for charging.

**Figure 6** shows the power consumption profile over a week from 100% level 2 charging stations (green) and 100% UFC stations (yellow). Due to the differences in the time to charge and the station power ratings, a comparison of the two scenarios shows that the peak and valley power parts of both profiles occur at different times of the day. It should be noted that within **Figure 6**, all other power profiles representing UFC replacement increments should fall between the 2 power curves.

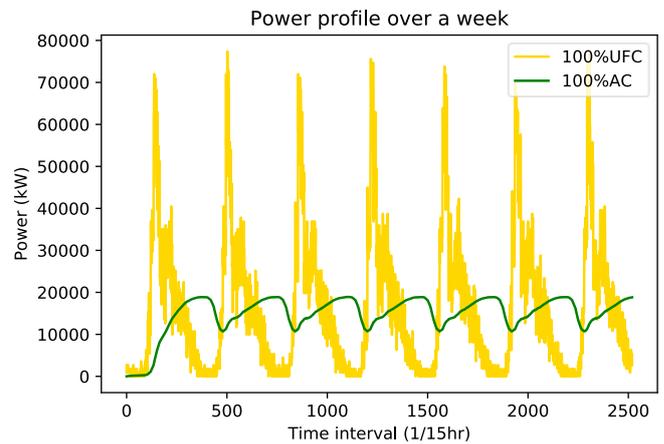

**Figure 6** – Weekly power profiles for 2 extreme scenarios

Subsequent to modeling the power curves for each UFC increment, the energy charge and demand charge are calculated according to PG&E's tariff at 15-min intervals and are applied to the EV user and station owner objective analyses. The peak power consumption is also used for the grid owner objective analysis. The results of these analyses are presented for one representative week in August and one representative week in January.

The objective for the station owner is to minimize the total cost, which is the sum of energy cost, demand cost and the O&M cost. **Figure 7** shows the total cost breakdown by these three categories from 0% UFC replacement to 100% UFC replacement computed at 10% increments. The energy cost is further broken down by the peak hour charge, part-peak hour charge and off-peak hour charge.

The conclusion of this analysis includes that the total cost increases as more AC charging stations are replaced by UFC charging stations. With respect to the summer week in **Figure 7(a)**, with increased UFC charging station replacement, the increase in total cost is primarily contributed by increase in the demand charge (brown). More power is consumed in peak and part peak hour periods (red and orange), however its contribution to total cost increase is negligible compared to the demand charge. The same trend and decomposition is observed for the winter week in **Figure 7(b)**. Meanwhile, since there's no peak hour charge in the winter, and by keeping the arrival profile same, the total cost in the winter is lower than in the summer. Thus, from the station owner's perspective, UFC replacement would be costly under current pricing scheme.

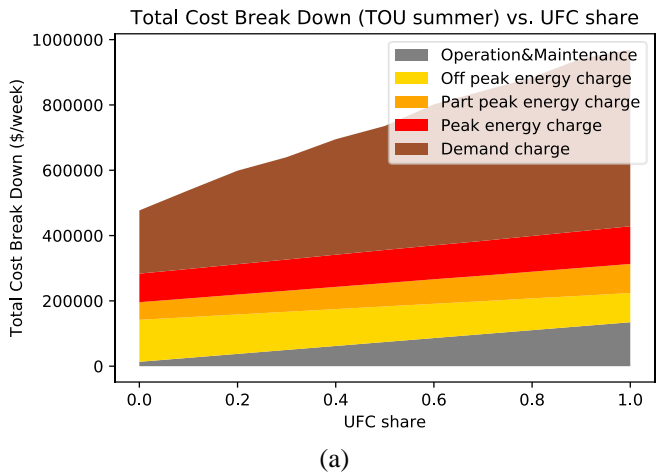

(a)

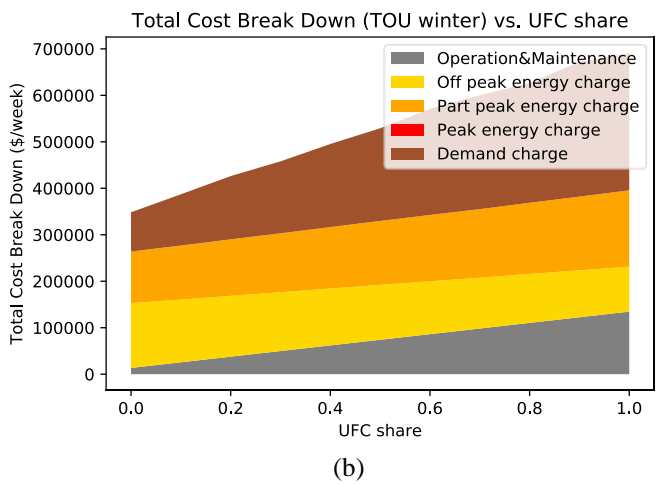

(b)

**Figure 7** – Weekly total charging breakdown vs. UFC share for (a) one week in August and (b) one week in January

The objective of the EV user is also to minimize the total electricity cost, while also minimizing the time to charge their EV. This is included in the total cost for the EV user, by assigning the VOT values mentioned previously to the time used for EV charging (20 hours for AC charging, 4 minutes for UFC charging).

Two representative assumptions of how people value their time are tested. In the first case, we think that waiting for EVs to fully charge deprive commuters of their work time. Thus, the minimum hourly wage under federal law of $7.25/hr. is applied [8]. However, because people do not wait in their EVs for 20 hours for a full-charged status, this may not be the best approximate for people's value of time. In the second case, it was found that the current EV charging schemes make drivers unable to perform work such as driving for Uber, showing that there is an opportunity cost associated with charging. Thus, the Uber driver's hourly pay is used as VOT factor, which is estimated at around $8.55/hr by Stephen Zoepf [7]. Additionally, a tipping point VOT factor is tested to see at which point would drivers value AC charging and UFC charging indifferently.

**Figure 8** presents the results of this analysis. and the tipping value of time level turns out to be relatively low: UFC charging would be preferred by driver if their valuation of time is higher than $0.70/hr in the summer, or $0.60/hr in the winter, which are both lower than the two previous assumptions we made about how people value their time. Since those two assumptions are seen as the lower-bound of people's value of time in the real world, from the EV user's perspective, it is justified to adopt using a UFC charging.

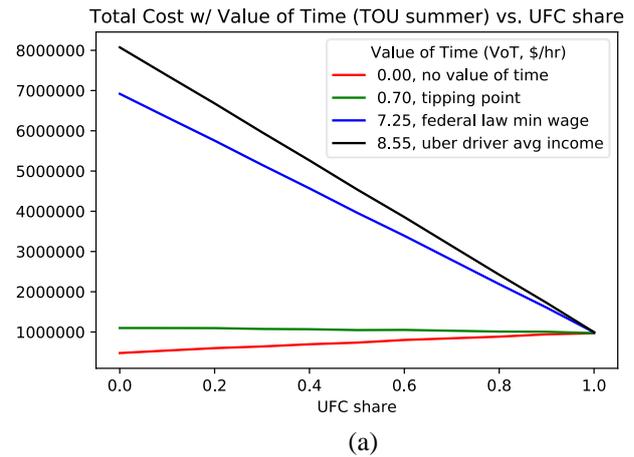

(a)

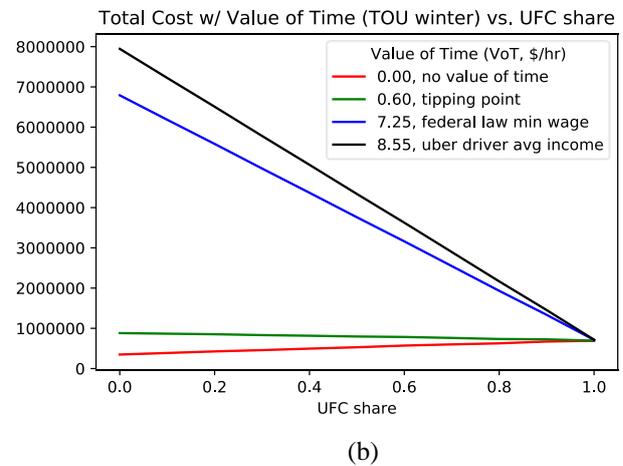

(b)

**Figure 8** – Impact of VOT on total cost vs. % UFC

The objective of grid operator is to minimize the observed peak load and the power consumed during the peak demand period, for the sake of grid stability and the generation constraints. **Figure 9a** shows the results of the aggregated energy consumption during the peak demand period. **Figure 9b** shows the observed peak load of the grid. Based on this analysis, both the aggregate energy consumption and the peak power increase from the base case of 100% level 2 charging stations in the bay. More specifically, from 100% level 2 charging station to 100% UFC stations, the aggregate energy consumed increases by about 38.2% and the observed peak load increases by 300%.

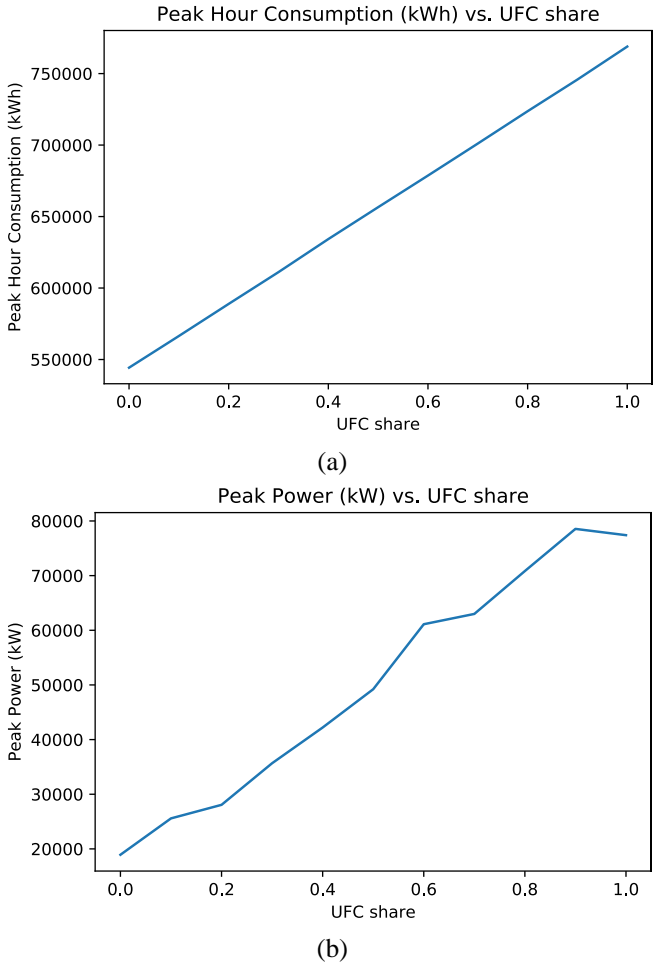

**Figure 9** – (a) Energy Consumed and (b) observed peak load vs. % UFC

According to **Figure 9(b),** a 100% UFC station system will contribute to an increase of 60MW (300%). In proportion to the peak load experienced by CA, this may be significant. The latest record of the CA peak load was 50,116 MW that occurred on September 1st, 2017 [10]. By performing a simple analysis where CA's peak load is multiplied by the proportion of the population in the Bay Area (7.76 million) and in CA (39.54 million), we find this peak load constitutes an increase of 0.6% of peak loading in the Bay area. The proportional analysis is given in **Equation 1**.

$$\text{Bay peak load} = \text{CA peak load} \times \frac{Bay\ population}{CA\ population} \quad (Eq.1)$$

This contribution of 0.6% potentially signifies the need of building new power plants, broadening transmission line, or, deploying battery storage to shave the observed peak load.

V. CONCLUSIONS AND FUTURE WORK

This project aims to build a framework for assessing the pros and cons associated with fully transitioning from level 2 EV charging stations to UFC stations within the San Francisco Bay Area. An analysis of how this new technology will affect the EV user, the charging station owner, and the grid operator was performed, and the following impacts were observed from this study:

1. **Power Consumption Impact:** The power consumption of the UFC charging station becomes sharper at a specific time period throughout the day and less smooth when compared to the level 2 charging station case.

2. **EV User Impact:** The proposed UFC transition increases the energy cost to the EV user, but when quantifying the cost of time waiting while charging the EV, transitioning to the UFC stations tends to be a more viable option.

3. **Station Owner Impact:** The station owner only takes into account the operational cost of using and maintaining the EV charging stations. This cost increases with the transitioning to the UFC stations, which makes this unfavorable to the station owner.

4. **Grid Operator Impact**: The aggregate energy consumption and the observed peak load both increase with the transitioning of the UFC stations. Thus, this transition is not favorable to this grid operator.

The first step for future work includes performing a power shaving analysis, coupled with battery storage, optimized for the objectives of each of the three stakeholders involved in this situation. The power shaving will be performed to optimize for cost (coupled with a VOT metric) and to optimize for aggregate energy and the observed peak load.

The second step of the analysis includes performing a sensitivity analysis for input parameters with notable uncertainty. These input parameters include existing traffic demand, proportion of EVs in traffic demand, daily UFC utilization, charging times, and cost of electricity over the time-of-use. Once the uncertainties in this analysis are understood, a further, refined assessment of the economic viability of implementing the UFC charging stations will be performed for each of the objectives defined.

Additional future work that can be performed to supplement this study includes performing the analysis of how PV solar – energy storage systems can supplement the transition from level 2 charging stations to UFC stations within locations with potential. This study can also be extended to analyse vehicle-to-grid (V2G) and vehicle-to-vehicle (V2V) charging modes, as well as understanding the charging behavior of EV commercial vehicles being charged through UFC stations.

Last, this EV framework can be used to study the market dynamics of transitioning from level 2 charging stations to UFC stations from different economical POVs. Different capital cost scenarios can be included in future within the current framework as the technology becomes cheaper as it is adopted more widely.

ACKNOWLEDGMENTS

This project is supported by CEE 272R Modern Power System Engineering. We are thankful to our project supervisor, Xiao (Mark) Chen, who provided expertise that greatly assisted the project. We are also grateful to Prof. Rajagopal and Teaching Assistants Sierra Gentry and Pepe Bolorinos for offering insightful classes and project instructions.